\newcommand{\FigureLength}{7.7cm}
\newcommand{\muHtwo}{\mu_{{\rm H}_2}}
\documentclass[showkeys,showpacs,prb,twocolumn,superbib]{revtex4}
\usepackage{graphicx}
\usepackage{color}

\begin{document}
\title{
First-principles study of graphene edge properties and flake shapes
}

\author{Chee Kwan Gan and David~J.~Srolovitz}
\affiliation{
Materials Theory and Simulation Laboratory, Institute of High
Performance Computing, 1 Fusionopolis Way, \#16-16 Connexis, Singapore
138632, Singapore
}
\date{25 Jan 2009}

\begin{abstract}

We perform density-functional theory calculations to determine the equilibrium shape of graphene flakes as a function of temperature and hydrogen partial pressure.  To do this, we first determine the edge orientation dependence of the edge energy and edge stress of graphene nanoribbons.  The edge energy is a monotonically increasing function of edge angle; increasing from the armchair orientation to the zigzag orientation.  However, reconstruction of the zigzag edge lowers its energy to less than that of the armchair edge in the absence of the hydrogen.  The edge stress for all edge orientations is compressive, however, reconstruction of the zigzag edge reduces this edge stress to near zero.  Hydrogen adsorption  is favorable for all edge orientations at sufficiently high hydrogen partial pressure; dramatically lowering all edge energies and all edge stresses.  It also lifts the reconstruction of the zigzag edge. Using these edge energy data within a grand canonical ensemble, we determine the equilibrium shape of a graphene flake to be hexagonal with armchair oriented edges at both very small and very large hydrogen chemical potential.  In the former case, the edges are hydrogen-free, while in the latter they are hydrogen terminated.  At intermediate hydrogen chemical potentials, graphene flakes show a more complex series of shapes, from zigzag edge terminated (near) hexagons, to dodecagons, to shapes with curved surfaces. Zigzag edge reconstruction produces graphene flakes with a six-fold symmetry, but with rounded edges. This shape is dominated by near zigzag edges. The compressive edge stresses will lead to edge buckling (out-of-the-plane of the graphene sheet) for all edge orientations, in the absence of hydrogen.  Exposing the graphene flake to hydrogen dramatically decreases the magnitude of the compressive edge stresses and reduces the buckling amplitude. 

\end{abstract}
\maketitle

\section{Introduction}
Graphene, a one-atom thick single sheet of $sp^{2}$-bonded carbon atoms in a honeycomb lattice, is a zero-gap semiconductor (semi-metal) that exhibits extraordinarily high electron mobility and shows considerable promise for applications in electronic and optical devices, high sensitivity gas detection, ultracapacitors and biodevices\cite{Novoselov04v306,Zhang05v438,Novoselov05v438,Son06v97,Schedin07v6,Geim07v6,Mohanty08v8,Stoller08v8,Echtermeyer08v29,CastroNeto09v81,Geim09v324}.  Graphene can be separated from bulk graphene using a peeling technique known as mechanical exfoliation\cite{Novoselov04v306}. Lithography techniques\cite{Li08v319} have recently been applied to fabricate patterned graphene nanoribbons (GNR) with widths below 100~\AA; such GNRs were used to produce graphene field effect transistors with on-off ratios as high as $\sim10^7$. There has been concerted efforts\cite{Meyer07v446,Wassmann08v101,Shenoy08v101,Jun08v78,Huang09v102,Girit09v323,Bets09v2} to investigate the stability of graphene sheets. Since applications of graphene necessarily employ sheets of finite extent, it is also of interest to inquire as to the equilibrium shape of such finite sheets or flakes.  

Two distinct definitions of shape arise in the context of graphene sheets. The first is associated with a finite, perfectly flat graphene
sheet. For such a sheet, the equilibrium shape is determined primarily by the energy of the graphene sheet edges.  Edge energies are related to the thermodynamically stable shape in exactly the same way that surface energies are used to predict the equilibrium shape of a three dimensional crystal through the Wulff construction~\cite{Pimpinelli98-book}. The second is associated with equilibrium elastic distortions of the graphene sheet as a result of edge stresses (analogous to surface stresses for a three-dimensional crystal). Several recent studies~\cite{Shenoy08v101,Jun08v78} have shown that the two highest symmetry graphene edges have compressive edge stresses, leading to elastic distortions of the regions near these sheet edges (edge warping).  In this paper, we employ density-functional theory (DFT) to determine graphene edge energies and edge stresses as a function of edge
orientation both in the absence and in the presence of hydrogen atoms.  We employ these results to investigate the equilibrium shape of a finite graphene sheet; both for the flat sheet (a two-dimensional intrinsic manifold) and one for which elastic distortions are permitted (a two-dimensional sheet embedded in three dimensions).

\section{Calculation Methods}
In order to determine the properties of the edges of graphene sheets, we employ both spin-polarized and non-spin-polarized DFT\cite{Hohenberg64v136,Kohn65v140} within the local density approximation using a self-consistent pseudopotential method, as implemented in the Siesta code\cite{Soler02v14}.  A double-$\zeta$ plus polarization basis set is used for the localized basis orbitals along with a large energy cutoff (400~Ry) for the real space integration. An electronic temperature of 0.01~eV is used to smear the Fermi-Dirac distribution.  In order to study graphene sheet edges, we focus on flat graphene nanoribbons (GNR) that are periodic in one direction and are separated by 16~\AA\ of vacuum in the other two directions. In the direction parallel to the ribbon axis, we use a $k$-point sampling criterion such that the number of $k$ points $n_k$ in the periodic direction is determined by the smallest integer that fulfills $n_k L = 136.96$~\AA, where $L$ is the length of the ribbon in the periodic direction. This corresponds to a 32-$k$-point scheme for $L=4.28$~\AA, the dimension of the smallest unit cell of an armchair edge.  We minimize the energy with respect to the positions of all atoms in the GNR plane; the relaxation is stopped when the force on each atom is less than 0.01~eV/\AA. We first determine the honeycomb lattice constant $a_0$ of the infinite graphene sheet, which is $a_0 = 2.471$~\AA.  This corresponds to a C-C bond length of $a_0/\sqrt{3} = 1.427$~\AA.  This agrees very well with the value of $a_0=2.468$~\AA\ found in another density-functional calculation~\cite{Reich02v66}.

\begin{figure}[htbp]
\centering\includegraphics[width=\FigureLength,clip]{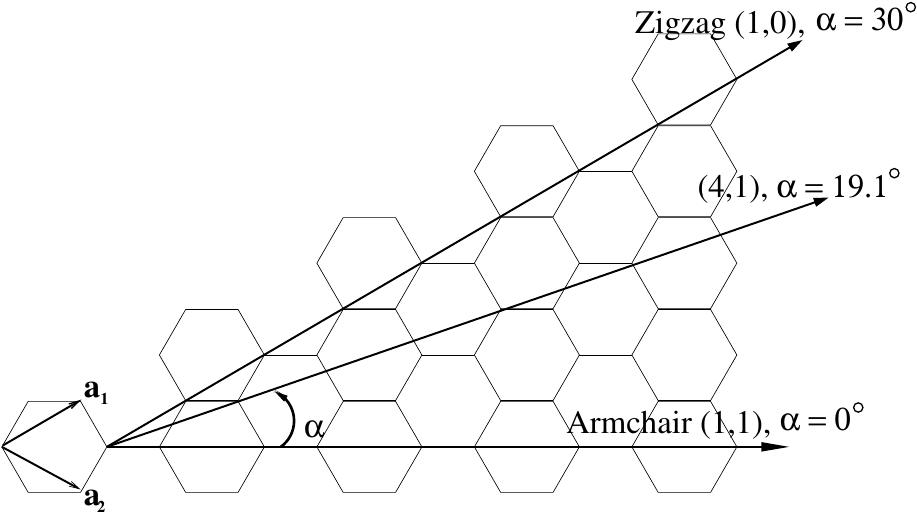}
\caption{A graphene  edge orientation is defined by the angle $\alpha$ between edge tangent vector ${\bf T}_e = n{\bf a}_1 + m{\bf a}_2$ and the armchair vector ${\bf a}_{ac} = {\bf a}_1 + {\bf a}_2$. The two highest symmetry edges are shown: the $\alpha = 0^{\circ}$ (1,1) armchair edge and the $\alpha=30^{\circ}$ (1,0) zigzag edge.  Also show is the more general $(4,1)$ graphene edge, corresponding to $\alpha= \cos^{-1}\left(\frac{5} {2\sqrt{7}}\right) = 19.1^{\circ}$. 
}
\label{fig:graphene-orientation.pdf}
\end{figure}

\begin{figure}[htbp]
\centering\includegraphics[width=\FigureLength,clip]{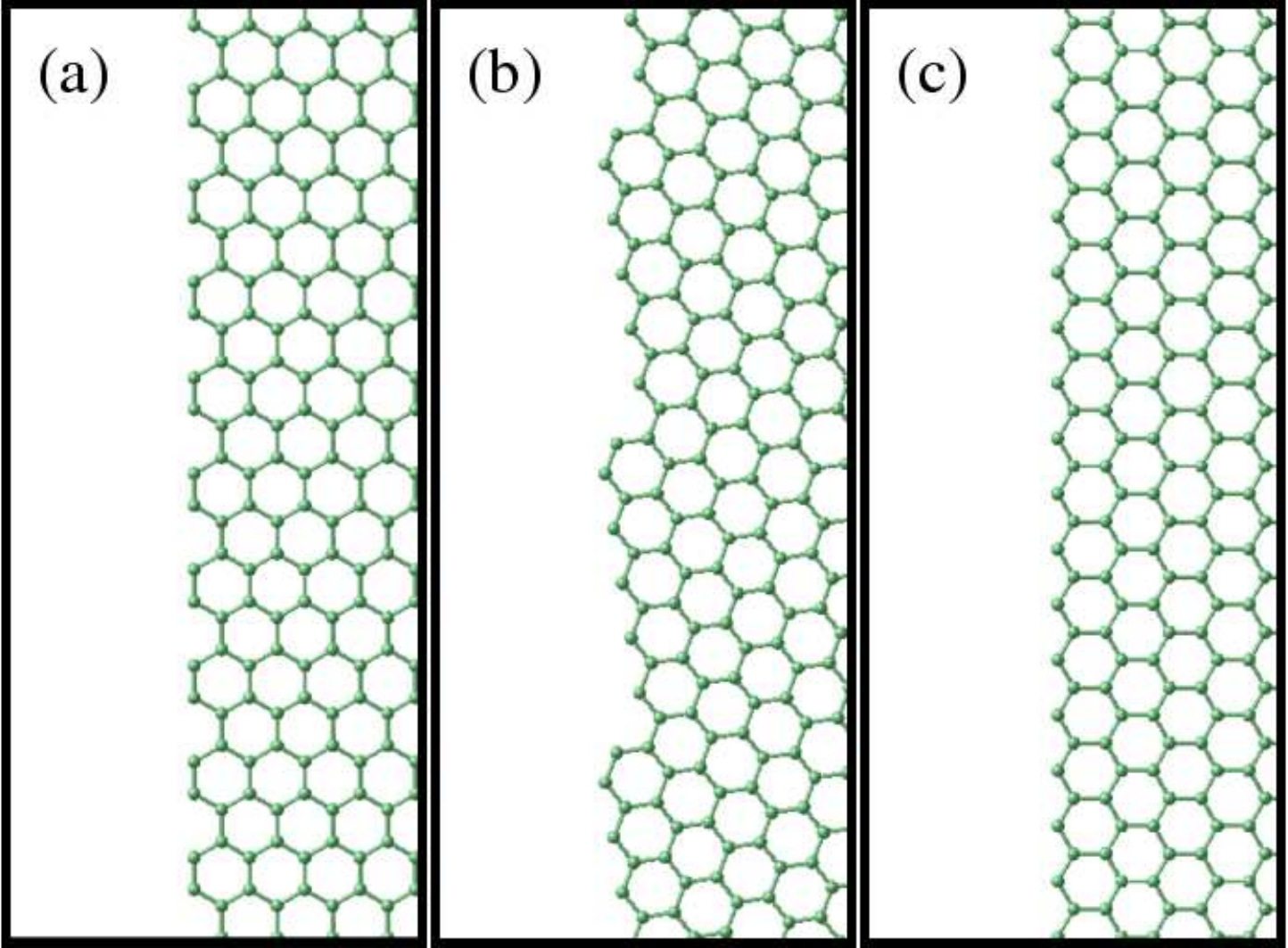}
\caption{(Color online) The edge structure of unreconstructed graphene sheets in the (a) armchair $\alpha=0^{\circ}$ (1,1), (b) $\alpha=21.05^{\circ}$ (5,1), and (c) zigzag $\alpha=30^{\circ}$ (1,0) orientations. 
}
\label{fig:ac-51-zz.pdf}
\end{figure}

A general graphene edge orientation can be described by a convention that is commonly used to describe the chirality of a single-wall carbon nanotube\cite{Saito98-book}.  We use $\alpha$ to describe the angle between the tangent to the edge ${\bf T}_e = n {\bf a}_1 + m {\bf a}_2$  and a second vector parallel to the armchair, as shown in Fig.~\ref{fig:graphene-orientation.pdf}.   This angle is given by $\alpha = \cos^{-1}\sqrt{\frac{3({n+m})^{2}}{4[({n+m})^{2}-mn]}}.$ Naturally, the armchair and zigzag edges correspond to $\alpha = 0^{\circ}$ and $30^{\circ}$, respectively.  Because of the symmetry of the honeycomb lattice, we only need to consider angles within the range of $ 0^{\circ} \le \alpha \le 30^{\circ}$.  In the present study, we study a discrete set of edge orientations in this $\alpha$-range:  $0^{\circ} (1,1)$, $6.59^{\circ} (3,2)$, $10.89^{\circ} (2,1)$, $16.10^{\circ} (3,1)$, $19.11^{\circ} (4,1)$, $21.05^{\circ} (5,1)$, and $30^{\circ} (1,0)$.  Figure~\ref{fig:ac-51-zz.pdf}b shows a $(5,1)$ graphene nanoribbon, corresponding to $\alpha = 21.05^{\circ}$; each repeat period of this edge structure contains 5 zigzag units (Fig.~\ref{fig:ac-51-zz.pdf}c) and a kink that has the armchair structure (Fig.~\ref{fig:ac-51-zz.pdf}a).

In the absence of hydrogen, the edge energy (per unit length) $E_{\rm edge}$ is 
\begin{equation}
E_{\rm edge} = \frac{1}{2L} [ E_{\rm GNR}(\alpha,w) - n_{\rm C} E_1],
\label{eq:EE}
\end{equation}
where $E_{\rm GNR}(\alpha,w)$ is the energy of an optimized GNR with edge orientation $\alpha$ and width $w$ that contains $n_{\rm C}$ carbon atoms, and $E_1$ is the energy per carbon atom in the infinite, flat graphene sheet with the lattice parameter that minimizes the energy. The factor of two in Eq.~\ref{eq:EE} accounts for the fact that a GNR has two edges.  In this work, we also consider the termination of all of the dangling bonds on the unreconstructed edges by hydrogen atoms.  (We only consider the case where all of the dangling bonds are saturated with a single H or no dangling bonds have H atoms to maintain the simplicity required to study many edge orientations.  We note that for ideal armchair and zigzag edges, more complex H-terminations have been considered\cite{Wassmann08v101}.)   In this case, the edge energy is 
\begin{equation}
E_{\rm edge+H}= \frac{1}{2L} \left[ E_{\rm GNR+H}(\alpha,w) - n_{\rm C} E_1 - \frac{n_{\rm H} E_{{\rm H}_{2}}}{2}  \right]
\label{eq:EdwithH}
\end{equation}
where $ E_{\rm GNR+H}(\alpha,w)$ is the energy of the graphene nanoribbon where the dangling bonds associated with carbon atoms on the edge are saturated with single H  atoms and $n_{\rm H}$ is the total number of such H atoms. $E_{{\rm H}_2}$ is the energy of the hydrogen molecule and the factor of $2$ in the last term accounts for the fact that each H$_{2}$ molecule only contributes one H  atom to each initially unsaturated bond. Note that Eq.~\ref{eq:EdwithH} reduces to Eq.~\ref{eq:EE} in the absence of hydrogen atoms, $n_{\rm H}=0$.

By analogy with the surface stress\cite{Gibbs-book1}, we define the edge stress $\tau_{\rm e}$ as the work $\Delta E_{\rm edge}$ required to 
stretch a graphene edge by an infinitesimal strain $\epsilon$ (parallel to the edge), $\Delta E_{\rm edge}/\epsilon$, or in differential form by 
\begin{equation}
\tau_{\rm e} =  \frac{dE_{\rm edge}}{d\epsilon}.
\label{eq:EdgeStress}
\end{equation}
(Note:  two definitions of surface stress $\tau_{s}$ have been widely employed in the literature:  (a) $\tau_{s}=d\gamma/d\epsilon$ and (b) $\tau_{s}=\gamma+d\gamma/d\epsilon$, where $\gamma$ is the surface energy.  We apply the former definition (a) here, i.e., $\tau_{e}=dE_{\rm edge}/d\epsilon$.  This is consistent with the method employed by Jun~\cite{Jun08v78}, but not by Huang {\it et al.}~\cite{Huang09v102} in their studies of graphene edges.  This definition yields the edge stress that produces elastic distortions of graphene nanoribbon edges~\cite{Shenoy08v101,Jun08v78}.) The derivative in Eq.~\ref{eq:EdgeStress} is determined using a method based upon the bulk stress calculation within the Siesta code~\cite{Soler02v14} that assumes an affine transformation of the atomic coordinates according to the applied strain.  Throughout this work, we adopt the convention that a negative (positive) edge stress corresponds to compression (tension).  

\section{Edge Energy}

\begin{figure}[htbp]
\centering\includegraphics[width=\FigureLength,clip]{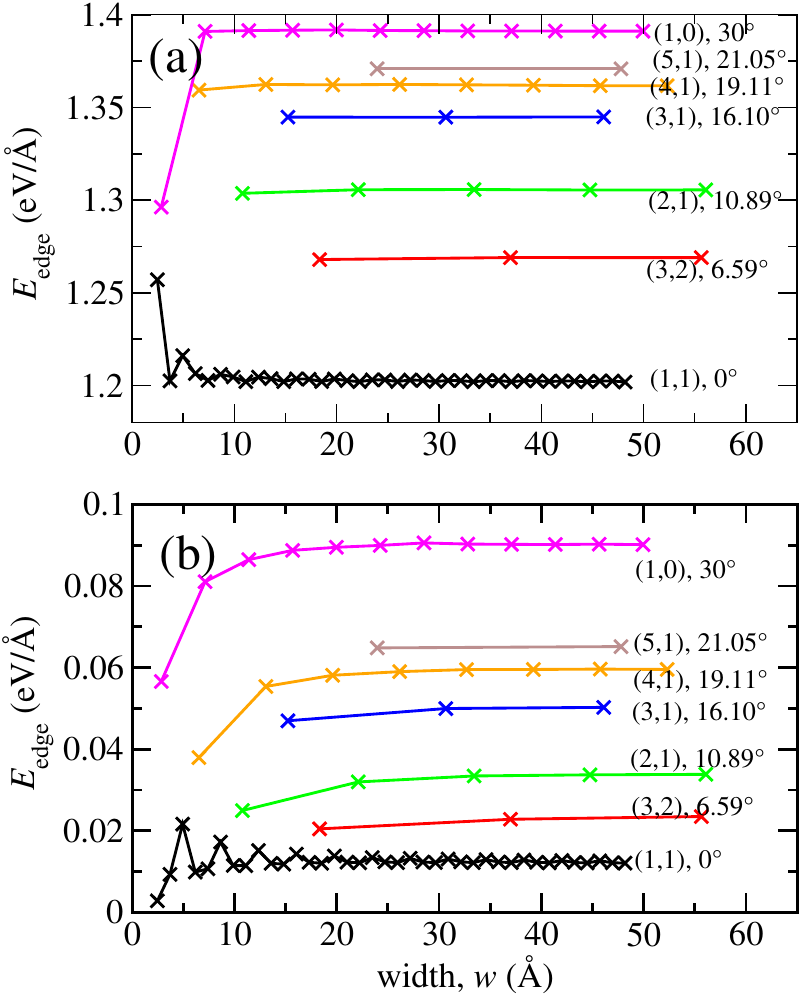}
\caption{(Color online) The edge energy as a function of GNR width of the unreconstructed graphene nanoribbon  (a) without hydrogen termination, and (b) with hydrogen termination for several GNR edge orientations.  
}
\label{fig:05-width-vs-edge-energy-all-direction.pdf}
\end{figure}

The first question we address, is how wide must a GNR be in order that the interactions between the opposite, parallel edges are negligible. Figure~\ref{fig:05-width-vs-edge-energy-all-direction.pdf} shows the edge energy versus GNR width, without and with H-termination, with spin-polarized calculations. The convergence of the edge energy in non-spin-polarized calculations show similar
behavior\cite{Gan09-unpublished}. (We note that earlier first-principles calculations have both included \cite{Son06v97,Huang09v102} and excluded\cite{Jun08v78,Koskinen08v101} spin polarization effects.) These figures clearly demonstrate  that the edge energy converges to a nearly width-independent value for $w \ge 30$~\AA\ for all GNR orientations, both without and with H-termination. 

\begin{figure}[htbp]
\centering\includegraphics[width=\FigureLength,clip]{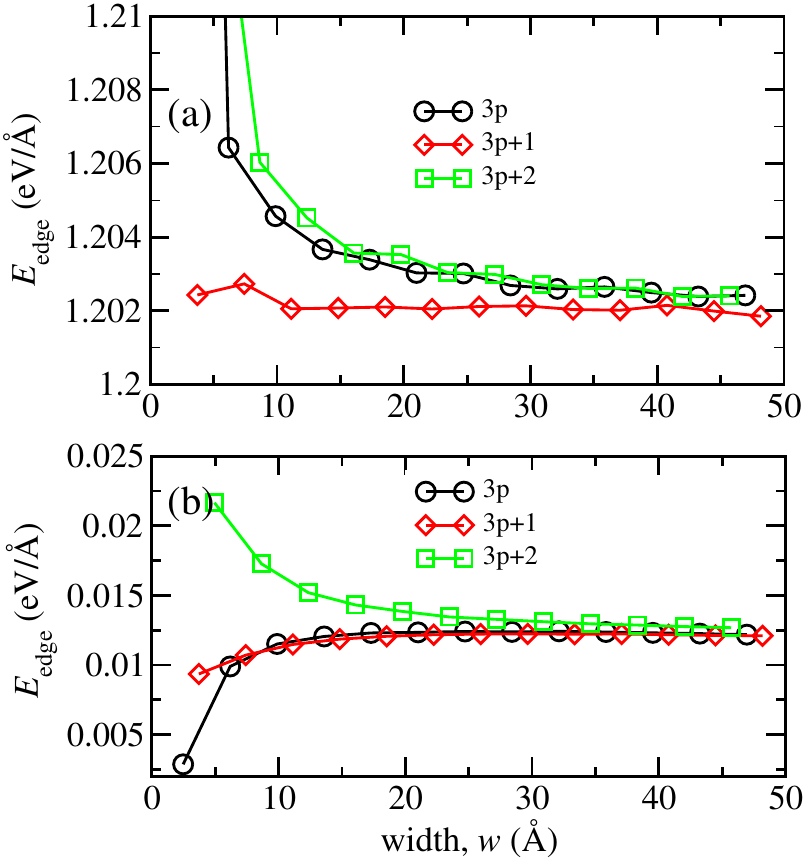}
\caption{
(Color online) The $\alpha=0^{\circ}$ (1,1) armchair edge as a function of ribbon width $w$ grouped according to the division of the data into three families (a) without hydrogen termination and (b) with hydrogen termination. 
}
\label{fig:04-AGNR-three-family-edge-energy.pdf}
\end{figure}

Unlike for all of the other edges, the $\alpha=0^{\circ}$ (1,1) armchair edge energy exhibits a decaying  oscillatory behavior with respect to GNR width $w$, as seen in Fig.~\ref{fig:05-width-vs-edge-energy-all-direction.pdf}.  This is consistent with earlier observations for the special case of an armchair GNR without H-termination\cite{Kawai00v62,Fujita97v66}.  Son {\it et al.}\cite{Son06v97} demonstrated that the energy bandgap of armchair GNRs can be grouped into 3 families according to the GNR width, $w$.  These families correspond to $ N_{a}=3p$, $3p+1$ and $3p+2$, where $p$ is an integer and $N_a$ is the number of carbon rows parallel to the armchair GNR axis.  Figures~\ref{fig:04-AGNR-three-family-edge-energy.pdf} a and b show the edge energy versus GNR width for the unterminated and H-terminated cases, where the data has been separated according to these families.   With this division, the armchair edge energy versus $w$ largely converges monotonically.

The converged edge energies  are shown in Figs.~\ref{fig:angle-gives-Eedge-all.pdf}a and b as a function of edge type, $\alpha$ and in Table~\ref{table:angle-edge-energy-edge-stress}. In non-spin-polarized calculations, the energy of the unreconstructed and non-H-terminated edges increases nearly linearly with $\alpha$; the armchair (zigzag) edge has the lowest (highest) edge energy, $1.202~$eV/\AA\ ($ 1.543$~eV/\AA). While the armchair edge energy is unaffected by spin polarization, the effect of spin polarization becomes increasingly important with increasing $\alpha$ (spin polarization always decreases edge energy). With or without spin polarization, the zigzag edge has a higher energy than the armchair edge (spin polarization reduces this difference by nearly a factor of two, from 0.341~eV/\AA\  to 0.189~eV/\AA ). The fact that the armchair has lower energy than the  zigzag edge can be understood  by the formation of triple bonds in the armrest regions of the armchair (as suggested by the fact that this bond length 1.258~\AA\ is much smaller that the bulk graphene C-C bond length, 1.427~\AA). In contrast, the zigzag edge does not have the opportunity to form triple bonds; the outer C-C bonds relax slightly to yield a bond length of 1.400~\AA. 

We have considered a low-energy, non-H-terminated reconstruction of the zigzag edge wherein two hexagons on the edge are transformed to a pentagon and a heptagon, called zigzag(57) reconstruction~\cite{Koskinen08v101}. Both non-spin-polarized and spin-polarized calculations
yield identical zigzag(57) edge energies, 1.147~eV/\AA, which are 0.055~eV/\AA\ lower than the $\alpha=0^\circ$ armchair edge. The same reconstructed edge is found to be 0.02~eV/\AA\ lower than the $\alpha=0^\circ$ edge in another DFT study\cite{Koskinen08v101}.

\section{Hydrogen Adsorption}

\begin{figure}[htbp]
\centering\includegraphics[width=\FigureLength,clip]{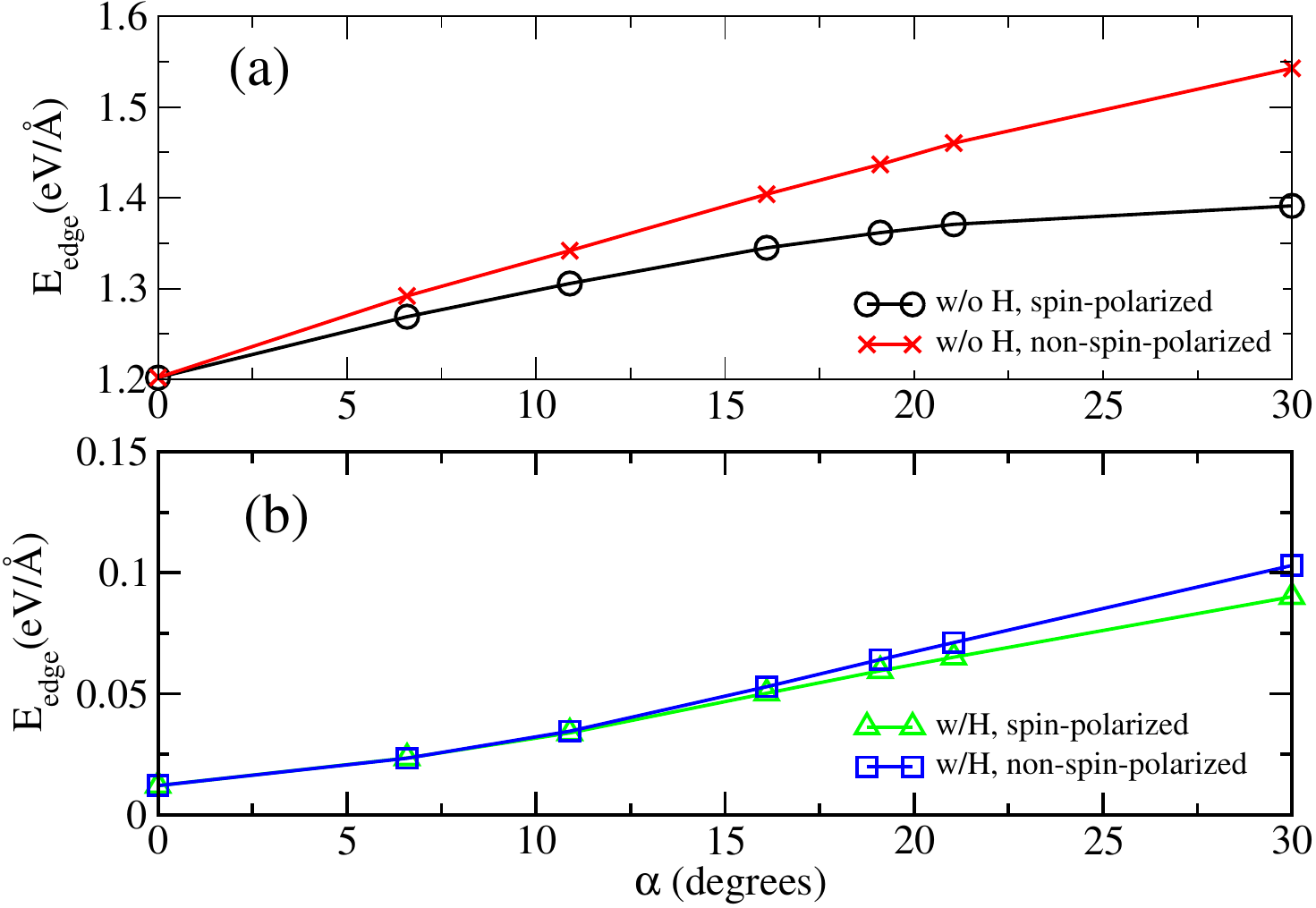}
\caption{ (Color online)
The graphene edge energy as a function of  the edge orientation $\alpha$  (a) without H-termination, and (b) with H-termination, with spin-polarized and non-spin-polarized calculations.
The edge energy for the non-H-terminated (H-terminated) is 1.147~eV/\AA\ (0.352~eV/\AA) reconstructed zigzag(57) edge using
spin-polarized calculations.
}
\label{fig:angle-gives-Eedge-all.pdf}
\end{figure}

When a single hydrogen terminates each of the dangling bonds along the ribbon edges, the unreconstructed edge energies drop by $1-2$ orders of magnitude, as seen in Figs.~\ref{fig:05-width-vs-edge-energy-all-direction.pdf}b and \ref{fig:angle-gives-Eedge-all.pdf}b. In non-spin-polarized calculations, the edge energy in the presence of hydrogen is maximum for the  zigzag [$\alpha=30^{\circ}$ (1,0)] edge ($E_{\rm edge} = 0.090$~eV/\AA) and minimum for the armchair [$\alpha=0$ (1,1)] edge ($E_{\rm edge} = 0.012$~eV/\AA). 
As in the non-H-terminated case, when the edges are H-terminated, spin polarization always decreases the edge energy.  The effect of spin-polarization is more important for edges with larger $\alpha$ and without H-termination.

In order to understand the effect of hydrogen on the GNR edges, we focus first on the energy of hydrogen adsorption
on the graphene edge $E_{\rm ads}$:  $E_{\rm ads} =\left(E_{\rm GNR} + n_{\rm H} E_{\rm H} - E_{\rm GNR+H}\right)/{n_{\rm H}}$. If we denote the number of H atoms adsorbed per unit length of the edge as $\eta_{\rm H} = n_{\rm H}/2L$, and the binding energy of the hydrogen molecule (relative to the isolated hydrogen atoms) as $E_{\rm b} = 2 E_{\rm H} -E_{{\rm H}_2}$, we can rewrite $E_{\rm edge+H}$ (see Eq.~\ref{eq:EdwithH}) as
\begin{equation}
E_{\rm edge+H}= E_{\rm edge} - \eta_{\rm H}( E_{\rm ads} - \frac{E_{\rm b}}{2}).
\label{eq:EdgeEnergyDiff}
\end{equation}
This change in edge energy (per unit length) upon saturation with hydrogen is simply the product of the number of H atoms adsorbed (per unit length) and the change in energy per hydrogen atom in going from H$_{2}$ in the gas to H on the edge.  For the unreconstructed edges, $\eta_{\rm H}$ has a very simple form:
\begin{equation}
\eta_{\rm H} = \frac{\cos\alpha}{a_0} \sqrt{\frac{4}{3}},
\end{equation}
where $a_0$ is the honeycomb lattice constant. $\eta_{\rm H}$ is $15.5\%$ larger for the armchair ($\alpha = 0^{\circ}$) than for the zigzag edge ($\alpha = 30^{\circ}$).

\begin{figure}[htbp]
\centering\includegraphics[width=\FigureLength,clip]{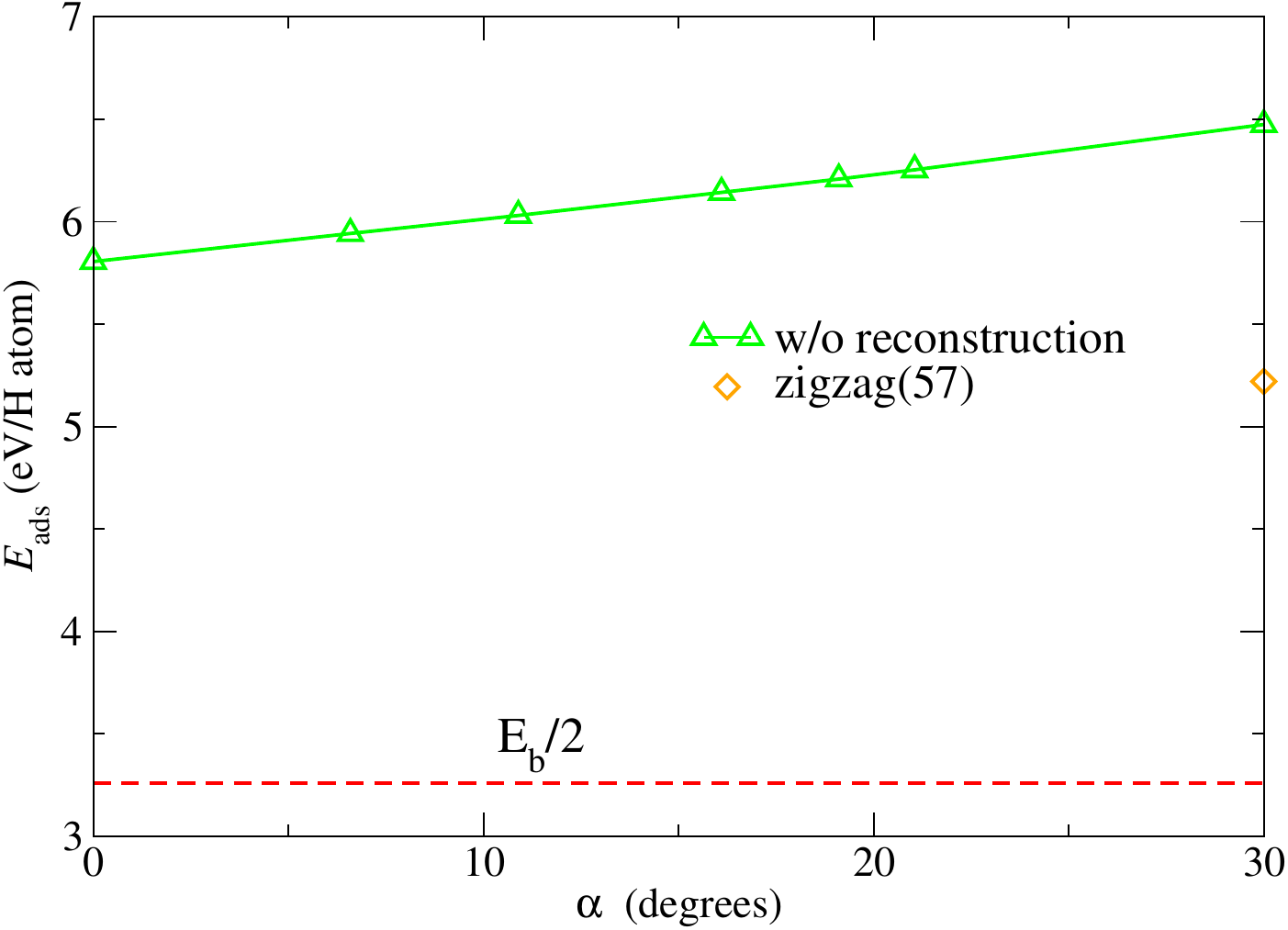}
\caption{(Color online) The hydrogen adsorption energy as a function of angle $\alpha$.  The binding energy of a hydrogen atom in H$_{2}$, $E_{b}/2$, is shown for reference.}
\label{fig:spin-polar-angle-gives-H-adsorption-energy.pdf}
\end{figure}

The main effect of adsorbing hydrogen on the edges is to decrease the edge energy by $\eta_{\rm H}(E_{\rm ads} - E_{\rm b}/2)$.  Figures~\ref{fig:angle-gives-Eedge-all.pdf}a and b show that the energy of the zigzag edge decreases (an energy decrease of 1.301~eV/\AA) upon hydrogen adsorption more so than does the energy of the armchair edge (an energy decrease of 1.190~eV/\AA).  This is contrary to what we should expect based upon the number of adsorbed hydrogen atoms per unit length, $\eta_{\rm H}$ (which decreases with increasing $\alpha$ as $\cos\alpha$).  However, the hydrogen adsorption energy $E_{\rm ads}$ increases with increasing $\alpha$ (as seen in Fig.~\ref{fig:spin-polar-angle-gives-H-adsorption-energy.pdf}) more quickly than $\eta_{\rm H}$ decreases.  This effect is associated with the fact that, before hydrogen termination, carbon atoms on the armchair edge have triple bonds to neighboring edge carbon atoms (and no dangling bonds), while the zigzag edge has no triple bonds (but does have dangling bonds)\cite{Koskinen08v101}. This suggests that hydrogen should bind more strongly to the zigzag edge atoms than to the armchair edge atoms. This implies $(E_{\rm GNR}-E_{\rm GNR+H})/n_{\rm H}$ will be larger for the zigzag edge than for the armchair edge.  Because the only term in $E_{\rm ads}$ that depends on edge type is $(E_{\rm GNR}-E_{\rm GNR+H})/n_{\rm H}$, this implies higher H adsorption energy to the zigzag edge than to the armchair edge.

The hydrogen adsorption energy on the zigzag(57) edge, 5.22~eV/atom, is significantly smaller than the adsorption energy on the unreconstructed zigzag edge (see Fig.~\ref{fig:spin-polar-angle-gives-H-adsorption-energy.pdf}).  This can be attributed to the fact that triple bonds formed in the heptagonal units on the zigzag(57) edge and to the lack of triple bonds on the unreconstructed zigzag edge. The difference in adsorption energy between the two zigzag edges explains why the hydrogenated zigzag(57) edge energy is (0.262~eV/\AA) higher than the hydrogenated unreconstructed zigzag edge (see Fig.~\ref{fig:angle-gives-Eedge-all.pdf}b). It is important to focus upon H adsorption energy $E_{\rm ads}$ for the zigzag and hydrogenated zigzag edges (presented above) {\it relative} to the binding energy of hydrogen in molecular hydrogen $E_{\rm b}/2$ (on a {\it per atom} basis).  This comparison shows that while reconstruction reduces the tendency for H-passivation of the zigzag edge, H will still bind to this edge.  (This conclusion differs from the  interpretation of the discussion in Ref.~\onlinecite{Koskinen08v101}.) 

\section{Edge Stress}

\begin{figure}[htbp]
\centering\includegraphics[width=\FigureLength,clip]{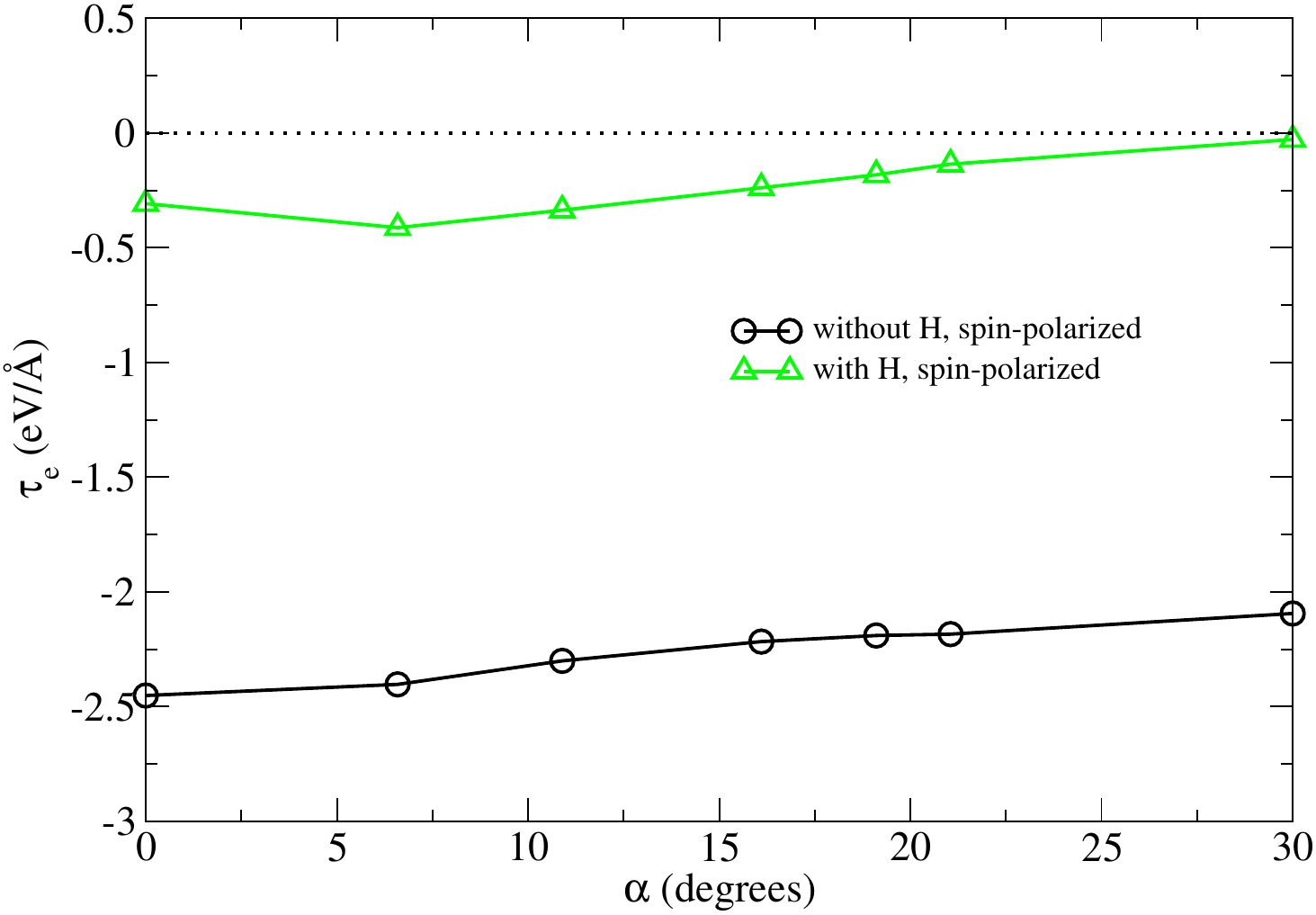}
\caption{ (Color online)
The edge stress as a function of edge orientation $\alpha$. The edge stress for the non-H-terminated (H-terminated) reconstructed zigzag(57) edge is $+0.14$~eV/\AA\ ($ +2.56$~eV/\AA).
}
\label{fig:angle-gives-tau-gamma-for-all-combinations.pdf}
\end{figure}

Since spin-polarization always lowers (or does not change) the edge energies, we focus on the edge stresses as determined using spin-polarized calculations.
The graphene  edge stresses, with and without hydrogen termination, are presented in Fig.~\ref{fig:angle-gives-tau-gamma-for-all-combinations.pdf}.  Without hydrogen termination, the edge stresses become more compressive with decreasing  $\alpha$. For the armchair [$\alpha=0^{\circ}$ (1,1)] edge $\tau_{\rm e} = -2.45$~eV/\AA\ and  for the zigzag [$\alpha=30^{\circ}$ (1,0)] edge $\tau_{\rm e} = -2.09 $~eV/\AA.  These values are of the same magnitude as those reported by Jun\cite{Jun08v78} (i.e.,  Jun reported  $\tau_{\rm e} =-2.640$~eV/\AA\ and $\tau_{\rm e} =-2.248$~eV/\AA\ for the armchair and zigzag edges, respectively).  Calculations of the edge stresses using an adaptive intermolecular reactive empirical bond order (AIREBO) potential, significantly underestimate the magnitude of the armchair edge stress $\tau_{\rm e}= -1.05 $~eV/\AA. 
However,  the AIREBO zigzag edge stress  $\tau_{\rm e}= -2.05$~eV/\AA\ agrees rather well with our DFT result $\tau_{\rm e}= -2.09$~eV/\AA.  While earlier studies only considered the edge stresses of the two highest symmetry edges, the results of  Fig.~\ref{fig:angle-gives-tau-gamma-for-all-combinations.pdf} show the edge stress for many intervening orientations.  Reconstruction of the zigzag edge to zigzag(57) structure greatly increases the zigzag edge stresses in the presence or absence of H-termination.  

Saturating the GNR edges with hydrogen significantly decreases the magnitude of the edge stresses (see  Fig.~\ref{fig:angle-gives-tau-gamma-for-all-combinations.pdf}).  Interestingly, saturating the zigzag GNR edge with hydrogen changes the edge stress  from highly compressive, $-2.09$~eV/\AA,  to nearly zero, $-0.03$~eV/\AA.  This is consistent with the results of an earlier study\cite{Jun08v78} that showed a significant drop in the magnitude of the edge stress upon saturating the edges with hydrogen. 
Finally we note that the compressive edge stress is caused by the interplay\cite{Bets09v2} between the change in the bond lengths, the accompanying changes in bond angles and the hybridization states of the atoms near the edges due to missing neighbors\cite{Groner06v780}. For example,  triple bonds form on armchair edges. This by itself, creates a tensile edge stress. However, the outermost edge atoms tend to move toward the interior of the graphene ribbon, increases the bond angles from the ideal $120^\circ$ to $\sim125^\circ$ (Refs.~\onlinecite{Bets09v2,Koskinen08v101}). This change in bond angle tries to force the atoms on the edge apart from one another, creating a compressive edge stress.


\section{Equilibrium Graphene Sheet Shape}

\begin{figure}[htbp]
\centering\includegraphics[width=\FigureLength,clip]{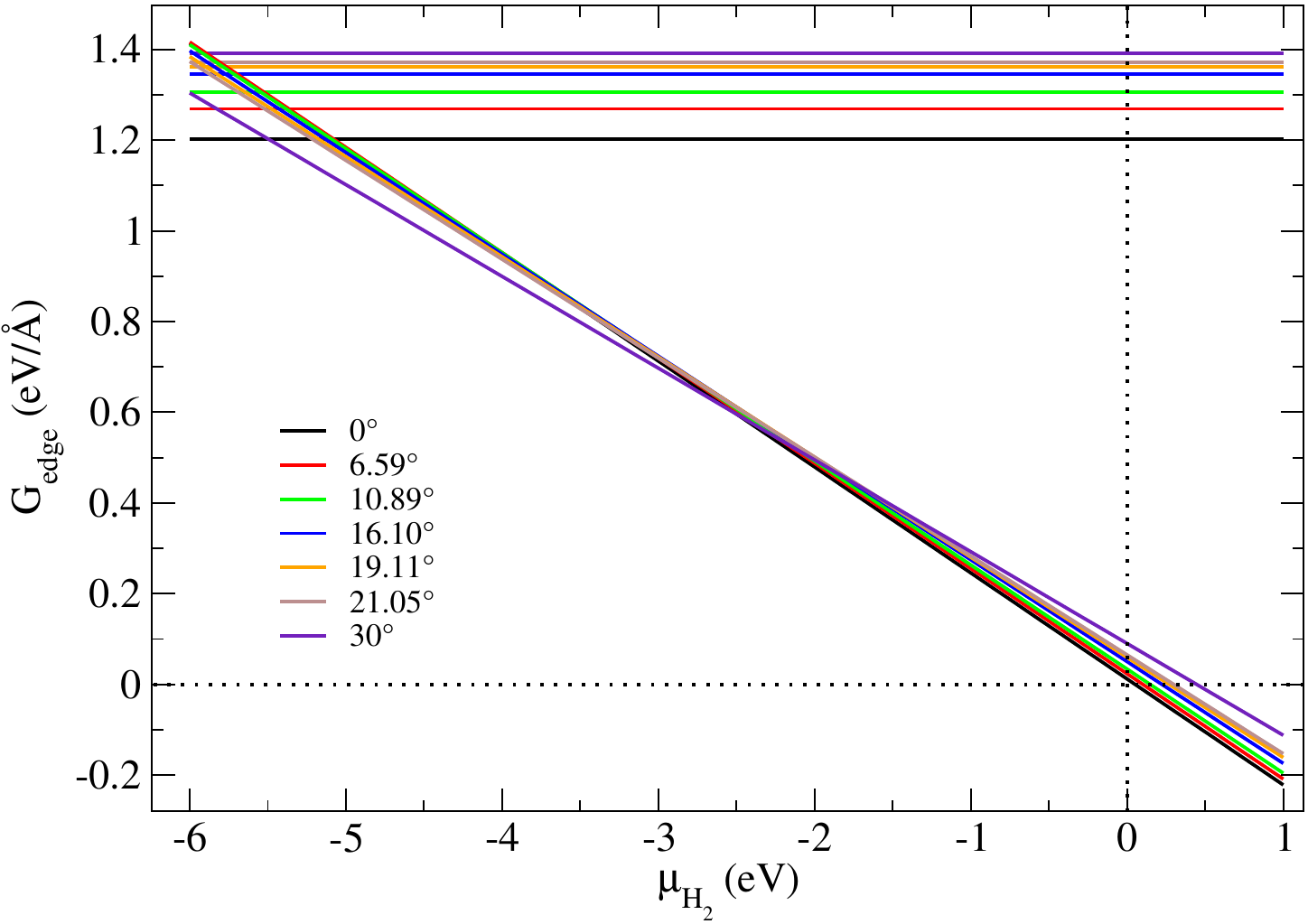}
\caption{(Color online) The Gibbs free energy difference $G_{\rm edge}$ as a function of the H$_2$ chemical
potential, $\muHtwo$.
The lines with
nonzero (zero) slopes correspond to edges with (without) hydrogen termination.
}
\label{fig:H2chemicalpot-gives-GibbsEnergy.pdf}
\end{figure}

In this section, we investigate the equilibrium shape of a flat graphene sheet, following a procedure akin to that used for determining the equilibrium shape of a 3-dimensional crystal via the Wulff construction (e.g., see Ref.~\onlinecite{Pimpinelli98-book}).  
The first step is to construct a polar plot of the edge energy $E_{\rm edge}$ as a function of the edge 
orientation $\alpha$. However, since the edge energy is a function of hydrogen density along the edge and this density depends on the hydrogen partial pressure and the temperature of the gas in which the finite graphene sheet exists, we focus on the edge free energy  $G_{\rm edge}$ as a function of the hydrogen chemical potential in the gas phase rather than the edge energy $E_{\rm edge}$.  We can write the edge free energy 
in the presence of a gas containing H$_2$ at a chemical potential $\muHtwo$ as\cite{Wassmann08v101} 
\begin{eqnarray}
G_{\rm edge} & = & \frac{1}{2L}\left[  E_{\rm GNR+H}(\alpha) - n_{\rm C} E_1 - \frac{n_{\rm H} (\muHtwo + E_{{\rm H}_2})}{2}  \right] \\
&=& E_{\rm edge+H} - \frac{\eta_{{\rm H}} \muHtwo}{2},
\end{eqnarray}
where the edge energy $E_{\rm edge+H}$ is given by Eq.~\ref{eq:EdwithH}. 
At temperature $T$ and H$_2$ partial pressure $P$, we have
\begin{equation}
\muHtwo = H^\circ (T) - H^\circ (0) - T S^\circ (T) + k_B T \ln \left( \frac{P}{P^\circ} \right),
\end{equation}
where $H^\circ $  and $S^\circ$ are the enthalpy and entropy of the gas at $P=P^\circ = 1$~bar. $H^\circ$ and $S^\circ$ may be obtained from, e.g., Ref.~\onlinecite{Chase98}. Using these expressions and the data in Table~\ref{table:angle-edge-energy-edge-stress},
we obtain the edge free energy $G_{\rm edge}$ as a function of $\muHtwo$
for the unreconstructed edges as shown in Fig~\ref{fig:H2chemicalpot-gives-GibbsEnergy.pdf}.

\begin{figure}[htbp]
\centering\includegraphics[width=\FigureLength,clip]{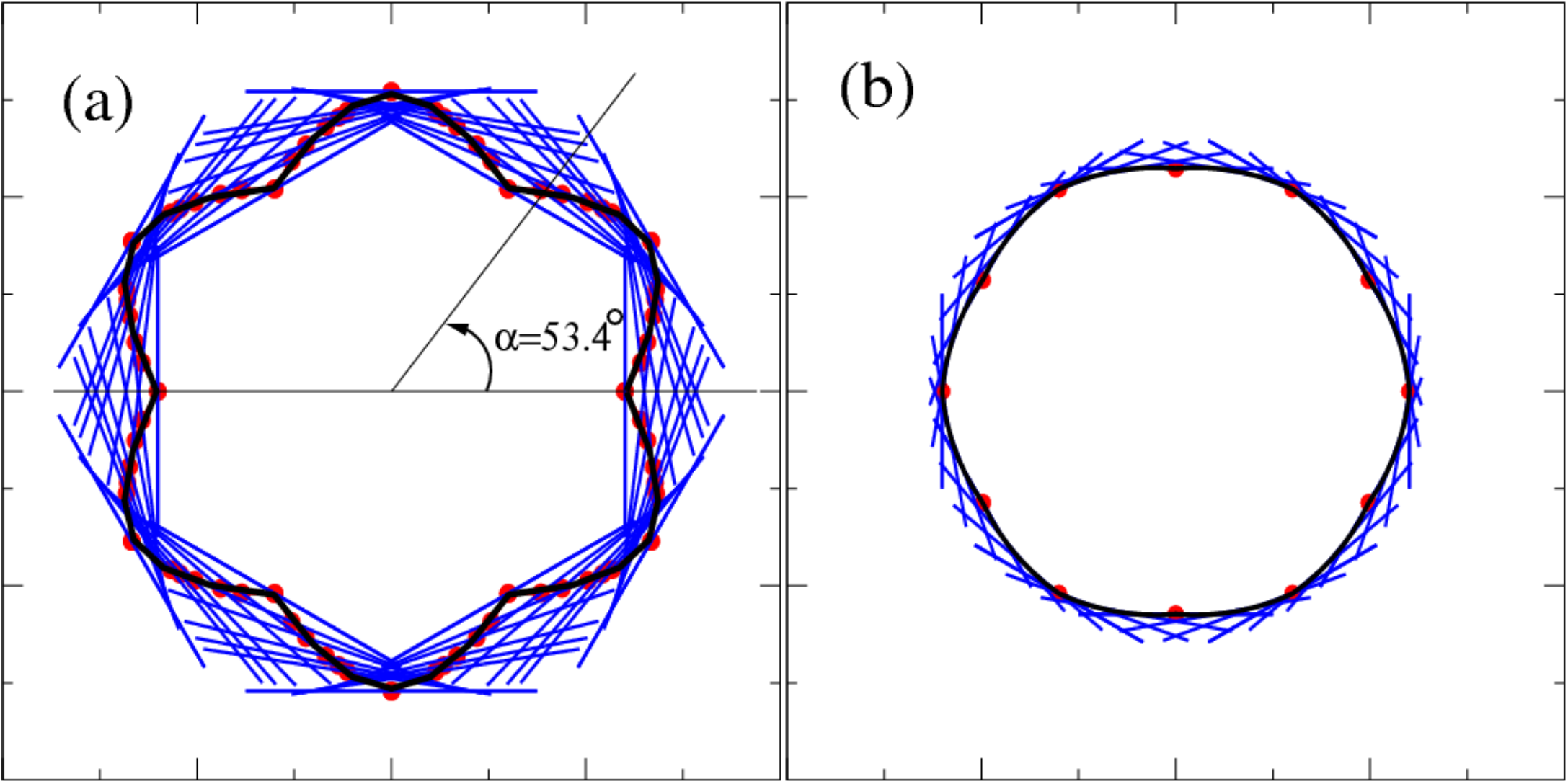}
\caption{(Color online) Polar plots of the unhydrogenated graphene edge free energy $G_{\rm edge}$ as a function of $\alpha$  (a) for unreconstructed edges and  (b) with reconstruction based upon a linear interpolation of the equilibrium armchair and zigzag(57) edge energies.  
The edge energy scales in (a) and (b) are identical. The red circles indicate data from the DFT calculations, the black lines are an interpolation of the data and the blue lines are drawn as lines perpendicular to the radius $G_{\rm edge}$ vectors.  The equilibrium, flat graphene flake shape is the inner envelope of the blue lines as obtained using the standard Wulff construction based on the edge energy curve.  
}
\label{fig:noH-noReconstruction.pdf}
\end{figure}

When $\muHtwo$ is large and negative, the edge free energies are lower in the unhydrogenated case than the free energies of the hydrogenated edges for all edge orientations.  In this case, we obtain a $G_{\rm edge}$ vs. $\alpha$ polar plot in Fig.~\ref{fig:noH-noReconstruction.pdf}a (we do not show the free energy of the hydrogen terminated edges here since they do not affect the Wulff shape under these conditions). Once we have a polar plot for $G_{\rm edge}$, we construct a ray from the origin at edge orientation angle $\alpha$ on this polar plot.  Next, we draw a line perpendicular to this ray at the point of its intersection with $G_{\rm edge}$.  This process is repeated for many (in principle, all) points around the polar plot.  The inner envelope of these lines represents the equilibrium (or Wulff) shape of the graphene flake.  This construction and the resultant equilibrium flat graphene flake shape (i.e., that which minimizes the edge energy at fixed flake area)  is shown in Fig.~\ref{fig:noH-noReconstruction.pdf}a for $\muHtwo = -7.0$~eV.  The equilibrium graphene flake shape is a hexagon with straight (1,1) armchair edges.

\begin{figure}[htbp]
\centering\includegraphics[width=\FigureLength,clip]{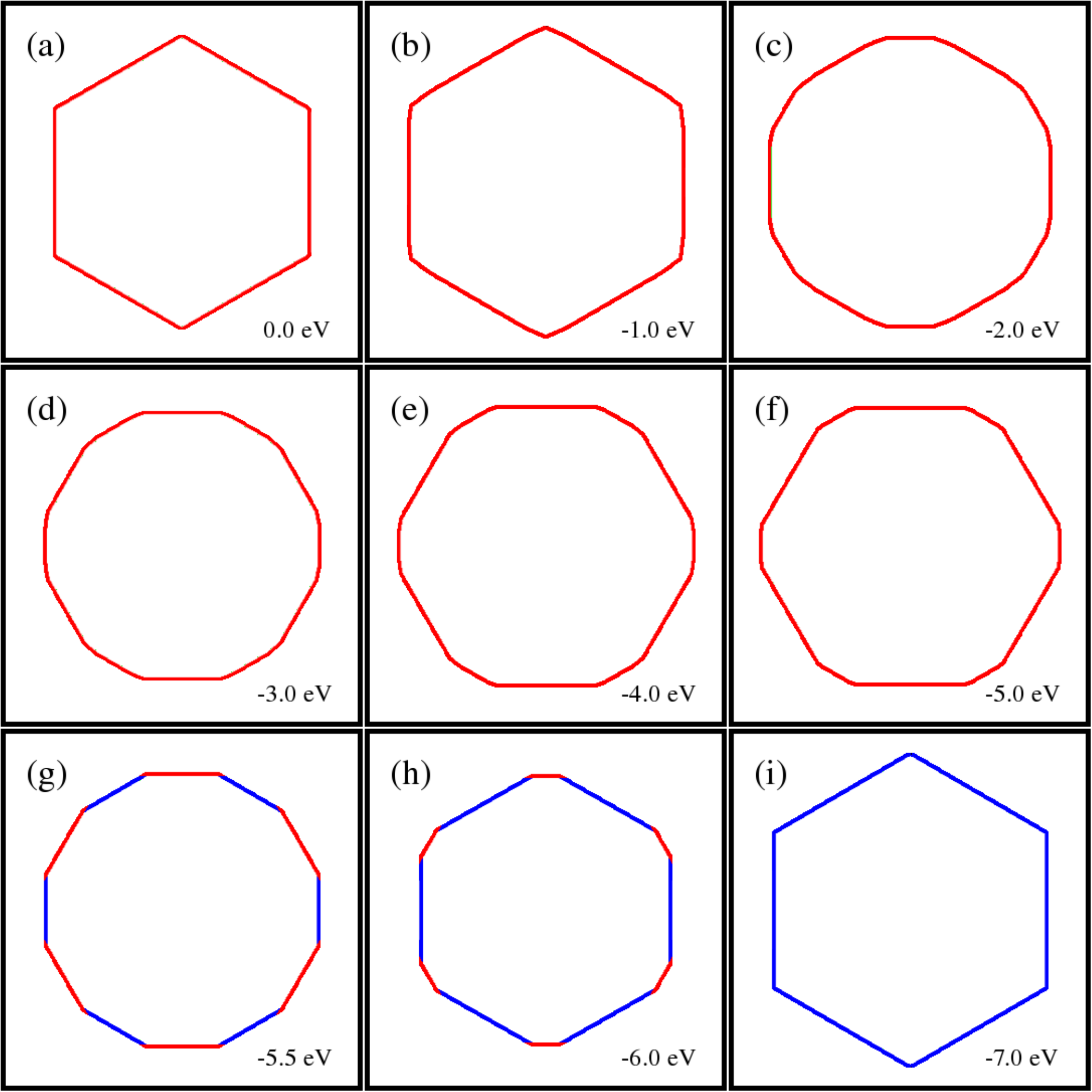}
\caption{(Color online) Equilibrium graphene flake shape as a function of 
H$_2$ chemical potential for $\mu_{\rm {H_2}}=$  (a) $0.0$ (b) $-1.0$ (c) $-2.0$, 
(d) $-3.0$ (e) $-4.0$ (f) $-5.0$ (g) $-5.5$, (h) $-6.0$ and (i) $-7.0$~eV. 
The red (blue) lines correspond cases with (without) hydrogen termination.
}
\label{fig:Wulff-plot-three-by-three-tile.pdf}
\end{figure}

Performing the Wulff construction for different values of $\muHtwo$, we determine the equilibrium graphene flake shape as a function of hydrogen partial pressure and temperature (as represented by $\muHtwo$).  These shapes are presented in Figs.~\ref{fig:Wulff-plot-three-by-three-tile.pdf}a-i for several values of $\muHtwo$ in the range $-7.0$~eV $\le$ $\muHtwo$ $\le 0.0$~eV. Figures~\ref{fig:Wulff-plot-three-by-three-tile.pdf}a and b  show that when $\muHtwo = 0.0 $ or  $-1.0$~eV, the equilibrium shapes have straight hydrogen-terminated armchair edges. When $\muHtwo=-2.0$~eV, the hydrogenated $30^\circ$ (zigzag) edge appears in the equilibrium shape. The hydrogenated zigzag edge becomes increasing dominant for $\muHtwo $ from $-3.0$ to $-5.0$~eV. Between $\muHtwo = -5.0$ and $-5.5$~eV, a new feature emerges in the equilibrium graphene flake shape, i.e.,  hydrogen-terminated armchair edges.  These edges grow in prominence with  decreasing $\muHtwo$. Finally, at $\muHtwo = -7.0$~eV, the  equilibrium shape is again a hexagon with straight armchair edges, as at $\muHtwo = 0.0$~eV.  However, for $\muHtwo = -7.0$~eV these edges are not hydrogenated while at $\muHtwo = 0.0$~eV they are.
 
The data presented in Fig.~\ref{fig:angle-gives-Eedge-all.pdf}a suggests that, in the absence of hydrogen, the zigzag edge reconstructs and that the reconstructed surface has lower energy than even the armchair edge (which does not reconstruct).  Since we do not have data for the reconstruction of all edges with $0^{\circ}<\alpha<30^{\circ}$, we make the simplifying assumption that the energies of these edges can be represented by a linear interpolation between the energies of the unreconstructed $\alpha=0^{\circ}$ armchair edge and that of the reconstructed $\alpha=30^{\circ}$ zigzag(57) edge.  The resultant polar  $G_{\rm edge}$ versus $\alpha$ plot is shown in Fig.~\ref{fig:noH-noReconstruction.pdf}b.  Unlike the polar  $G_{\rm edge}$ versus $\alpha$ plot for the unreconstructed case, this one is relatively smooth and nearly circular, yet a six-fold symmetry is obvious.  Performing the Wulff construction on this interpolated  $G_{\rm edge}$ versus $\alpha$ plot produces an equilibrium graphene flake shape that has six-fold symmetry, but no flat surfaces.  The absence of flat surfaces or facets can be traced to the fact that the polar  $G_{\rm edge}$ versus $\alpha$ plot here shows no sharp cusps (discontinuity of slope).  More interesting, however, is that when we allow edge reconstruction to occur (in the absence of hydrogen), the resultant six-fold equilibrium shape is rotated relative to that without edge reconstruction.  That is, the equilibrium graphene flake shape is dominated by surfaces at or near the zigzag edge orientation.


The above analysis focused on the equilibrium shape of a flat, finite graphene sheet (a two-dimensional intrinsic manifold).  However, a flat graphene flake is unstable against thermal rippling at finite temperature\cite{Barnard08v128}. Compressive edge stresses can also lead to elastic distortions of/near the graphene flake/ribbon edges, as discussed in the literature\cite{Meyer07v446,Shenoy08v101,Fasolino07v6}.  These distortions may be described as the buckling of the edges, associated with the compressive edge stress but constrained by the elasticity of the interior of the graphene flake.  The amplitude of the buckling $A$ has been described\cite{Shenoy08v101} as

\begin{equation}
A = \left( \frac{-\lambda \tau_{\rm e}}{( \pi\sqrt{20 + 14 \sqrt{7}}/18 )M}  \right)^{1/2}
\label{eq:BucklingAmplitude}
\end{equation}
where $\lambda$ is the wavelength of the buckling, $M = Y_b/(1-\nu^2)$, $Y_{b}$ and $\nu$ are the Young's modulus and the Poisson ratio of the graphene sheet,  and where we have neglected the relatively minor effect of the elastic stiffness of the edge.  Assuming a typical wavelength of $\lambda=10$~nm and $M=2000$ eV/nm$^{2}$  for graphene\cite{Shenoy08v101}, we can use the data presented above to determine the buckling amplitude $A$ as a function of edge orientation $\alpha$ and examine the effect of hydrogenation of the edge.

\begin{figure}[htbp]
\centering\includegraphics[width=\FigureLength,clip]{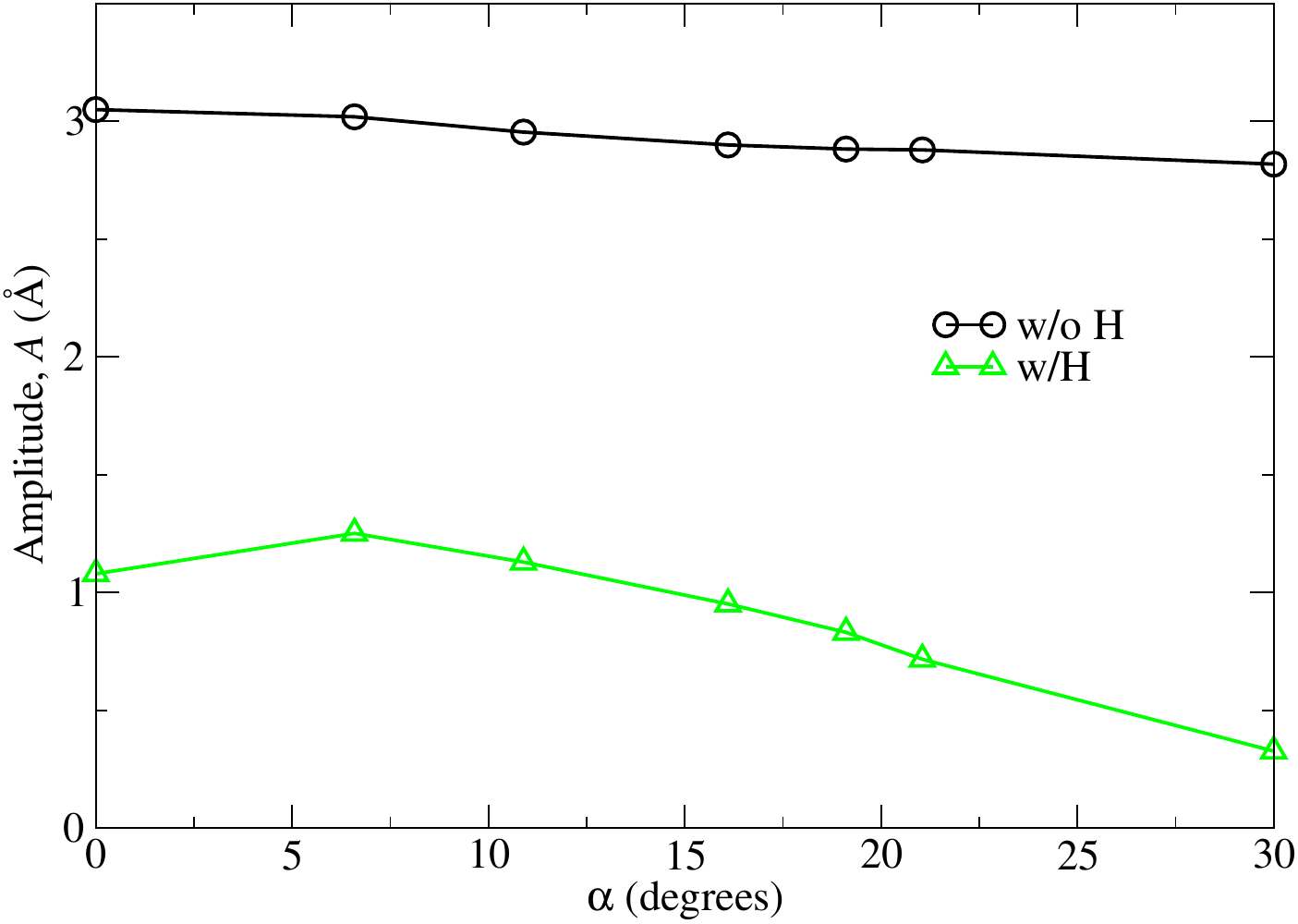}
\caption{(Color online) The edge buckling amplitude for the pure and hydrogenated, unreconstructed edges, as per Eq.~\ref{eq:BucklingAmplitude} using the data from Fig.~\ref{fig:angle-gives-tau-gamma-for-all-combinations.pdf}.  
}
\label{fig:BucklingAmplitude}
\end{figure}

Figure~\ref{fig:BucklingAmplitude} shows the buckling amplitude from Eq.~(\ref{eq:BucklingAmplitude}) using the edge stress data from Fig.~\ref{fig:angle-gives-tau-gamma-for-all-combinations.pdf} as a function of $\alpha$, with and without hydrogen on the edges.  The edge buckling amplitude found here is 3.05 and 2.82~\AA\ for the unreconstructed armchair and zigzag edges without hydrogen, respectively, which is roughly consistent with the magnitude of the buckling amplitude reported in Ref.~\onlinecite{Shenoy08v101}.  
The buckling amplitude varies from 2.82 to 3.05~\AA\ over the entire range of $\alpha$ for the  unreconstructed edges without hydrogen.  
Interestingly, if the zigzag edge reconstructs to the zigzag(57) structure, the buckling amplitude, as per Eq.~(\ref{eq:BucklingAmplitude}), is imaginary (the edge stress is positive);  implying that the reconstructed zigzag edge does not buckle. 

When the edges are saturated with hydrogen, the edge stress drops dramatically (by a factor of $\sim$ 7.9 to 70 for the armchair and zigzag edges, respectively) and the buckling amplitude drops to  $1.08$ and  $0.33$~\AA\ for the armchair and zigzag edges (see Fig.~\ref{fig:BucklingAmplitude}).  This represents a drop in edge buckling amplitude by a factor of $\sim$ 2.8 to 8.5. In short, edge hydrogenation strongly reduces the edge stress driven graphene edge buckling to the point that it is, for most purposes, negligible.

\section{Conclusions}

We have employed spin-polarized and non-spin-polarized density-functional calculations of graphene nanoribbons to determine the edge orientation dependence of the edge energy and edge stress (akin to surface stress in three-dimensional materials). 
We found that the spin-polarized calculations always give edge energies that are lower than (or the same as)
those obtained in non-spin-polarized calculations. The effect of spin-polarization is important for edges with larger $\alpha$ and for edges without hydrogen termination.
In the absence of reconstruction, the edge energy is a monotonically increasing function of 
the edge orientation angle (increasing from the armchair orientation to the zigzag orientation) for both
H-terminated and unhydrogenated edges.  Reconstruction of the zigzag edge, however, lowers its energy to less than that of the armchair edge.  The edge stress for all edge orientations is compressive.  Reconstruction of the zigzag edge reduces the edge stress to near zero (very slightly tensile).  Hydrogen adsorption to graphene edges is favorable for all edge orientations, with a larger adsorption energy to the unreconstructed zigzag edge than to the armchair.  Hydrogen adsorption dramatically lowers all edge energies and all edge stresses.  It also lifts the reconstruction of the zigzag edge. 
The thermodynamic properties of graphene edges play a key role in determining the equilibrium shape of a finite graphene sheet or flake.  
Using the grand canonical ensemble approach, we were able to account for the role of hydrogen partial pressure and temperature within the gas phase.  Using this approach, coupled with the Wulff construction, we determined the equilibrium shape of (unreconstructed) graphene flakes as a function of the hydrogen chemical potential.  We find that equilibrium graphene flakes are hexagonal with straight armchair edges for very small ($\muHtwo \lesssim -7.0 $~eV) or very large ($\muHtwo \gtrsim 0$) hydrogen chemical potentials. For intermediate values of the hydrogen chemical potential, the equilibrium graphene flake shape can be (nearly) hexagonal with zigzag facets, have rounded edges, or up to twelve facets.  However, if the zigzag edges reconstruct (as is thermodynamically favored), graphene flakes will have a six-fold symmetry, but with rounded (rather than faceted) edges. Interestingly, in this case, the six-fold symmetry is rotated with respect to the hydrogenated case and will be dominated by near zigzag edges. 

The compressive edge stresses will lead to edge buckling (out-of-the-plane of the graphene sheet) for all edge orientations, in the absence of hydrogen (or large negative hydrogen chemical potential).  The edge buckling amplitude is expected to be approximately 3 \AA, but this value depends on buckling wavelength.  Exposing the graphene flake to hydrogen will dramatically decrease the buckling amplitude to the point that it may either disappear or be too small to be of any significance.

\section{Acknowledgment}
The authors gratefully acknowledge useful discussions with Paulo Branicio, Yong~Wei Zhang, and Nathaniel Ng.  The basis sets and pseudopotentials were provided by Young-Woo Son and Su Ying Quek.  We are also grateful to Julian Gale and Su Ying Quek for their input on the use of the Siesta code. 
\bibliographystyle{apsrev}

\newpage
\pagebreak
\newpage



\pagebreak
\pagebreak
\newpage

\begin{table}[p]
\begin{center}
\begin{tabular}{c|l|r|l|r}
\hline\hline
& \multicolumn{2}{c|}{$E_{\rm edge} $ ($\tau_e$) }& \multicolumn{2}{c}{$E_{\rm edge+H} $ ($\tau_e$) }  \\
\hline
$\alpha$, $(n,m)$ & SP & NSP & SP & NSP \\
\hline
\ $0.00^{\circ}$, $(1,1)$ & 1.202 (-2.45) & 1.202 (-2.45) & 0.012 (-0.31) & 0.012 (-0.31)  \\
\ $6.59^{\circ}$, $(3,2)$ & 1.269 (-2.40) & 1.292 (-2.15) & 0.024 (-0.41) & 0.023 (-0.40)  \\
$10.89^{\circ}$, $(2,1)$ & 1.306 (-2.30) & 1.342 (-1.92) & 0.034 (-0.34) & 0.035 (-0.33) \\
$16.10^{\circ}$, $(3,1)$ & 1.345 (-2.22) & 1.404 (-1.94) & 0.050 (-0.24) & 0.053 (-0.22)  \\
$19.11^{\circ}$, $(4,1)$ & 1.362 (-2.19) & 1.437 (-2.01) & 0.060 (-0.18) & 0.064 (-0.16)  \\
$21.05^{\circ}$, $(5,1)$ & 1.371 (-2.18) & 1.460 (-2.06) & 0.065 (-0.14) & 0.071 (-0.12)  \\
$30.00^{\circ}$, $(1,0)$ & 1.391 (-2.09) & 1.543 (-2.55) & 0.090 (-0.03) & 0.103 (-0.01)  \\
\hline
\end{tabular}
\end{center}
\caption{The edge energy (in eV/\AA) and edge stress (enclosed by
parenthesis, in eV/\AA) for
spin-polarized (SP) and non-spin-polarized (NSP) calculations, both without ($E_{\rm edge}$) and with ($E_{\rm edge + H}$)
hydrogen terminations.
}
\label{table:angle-edge-energy-edge-stress}
\end{table}

\end{document}